# The reproductive number of Zika in municipalities of Antioquia, Colombia: stratifying the potential transmission of an ongoing epidemic

Juan Ospina[1], Doracelly Hincapie-Palacio[2], Jesús Ochoa[2], Adriana Molina[2], Guillermo Rúa[3], Dubán Pájaro[4], Marcela Arrubla[4], Rita Almanza[5], Marlio Paredes[6], Anuj Mubayi[6]

[1]Eafit University, Medellín, Colombia

[2]National School of Public Health, 'Héctor Abad Gómez', Epidemiology Group, University of Antioquia, Medellín, Colombia

[3]Group of Entomology, Faculty of Medicine, University of Antioquia, Medellín, Colombia

[4] Secretary Regional Health and Social Protection of Antioquia. Medellin Colombia

[5]Secretary of Health. Medellin Colombia

[6] Mathematical Theoretical Biology Institute; School of Human Evolution and Social Change; Simon A. Levin Mathematical Computational Modeling Science Center; Arizona State University, Tempe, Arizona

**Summary**

Introduction: Zika epidemic in America was declared a public health emergency of international concern after the rapid spread in the region. Stratification of the potential transmission of the disease is needed to address the efforts surveillance and disease control. The goal of this research is to compare the basic reproductive number of Zika in different municipalities, from an SIR model with implicit vector dynamics, based on the daily case reporting data of Antioquia, Colombia, the second most affected country after Brazil.

Methods: An simple SIR model with implicit vector dynamics was derived and used. The approximate solution of the model in terms of individuals recovered in each time unit, allowed to obtain estimate of the model parameters including the basic reproduction number (Ro) and its 95% confidence intervals. These parameters were estimated via fitting the solution of the model to the daily reported cumulative cases data from the regional surveillance system.

Results: Ro was estimated for 20 municipalities, all located at less than 2200 meters above sea level. From January to April 2016, between 17 and 347 cases were reported from these municipalities. Of these, 15 municipalities had a high potential for transmission (Ro>1) and 5 had less potential (Ro<1), although in 3 of these, transmission was later possible because its upper confidence interval of Ro was greater than one.

Conclusion: Surveillance and control of Zika should be directed primarily to municipalities with Ro>1. Furthermore, strategies that will strengthen the detection and management of cases in the remaining municipalities should be beneficial to the whole Antioquia region.

## Introduction

In May 2015, the Pan American Health Organization (PAHO) / World Health Organization (WHO) issued an epidemiological alert for the epidemic Zika in the Americas, after confirmation of autochthonous transmission in Brazil. Later on February 1, 2016, WHO declared the Zika epidemic in the Americas as a Public Health Emergency of International Concern [1, 2].

The Zika virus is an arbovirus of the genus *Flavivirus* (family Flaviviridae) transmitted by the *Aedes aegypti* mosquito, also vector of dengue and chikungunya[3]. Zika is considered a public health problem in America, by the rapid spread compared with previous epidemics in Africa and Asia and the high occurrence of severe neurological complications (Guillain-Barré syndrome) and occurrence of congenital syndromes related to Zika infection[4].

Forty countries / territories Americas reported autochthonous transmission from week 17 to week 25 from 2015 - 2016. Until July 7/2016, Brazil reported 78% of confirmed cases (64 311/81 914), 38.7 % of suspected cases (161 241/416 958) and 1,656 cases of congenital syndrome with suggestive evidence of congenital infection and 255 laboratory confirmed. Colombia was the second country to report 10.4% (n = 8506) of confirmed cases, 21.1% of suspected cases (n = 87,844), 11 confirmed cases of microcephaly related to Zika and 102 cases under study. Next in order, Puerto Rico (2.6% of confirmed cases) and Venezuela (2% of confirmed cases and 11.8% of suspected cases).[5]

PAHO / WHO on alert issued, recommended countries to establish integrated control to reduce virus transmission measures, including identifying areas of high risk of transmission[2]. This study estimated the basic reproductive number ($R_0$) as an indicator of potential Zika transmission in Antioquia. $R_0$ is the average secondary cases generated by an infecting when introduced into a completely susceptible population.[6]

Antioquia is located in northwestern Colombia with more than 6 million people. The territorial division consists of 125 municipalities divided into 9 subregions (Figure 1). It has great diversity of tropical climate, with the presence of *A. aegypti*, distributed in more than 80% of its land area. Antioquia reported the highest number of municipalities with laboratory-confirmed at week 28 of 2016[7]. Antioquia had a rapid spread of the disease in some subregions, cases presented more aggressive symptoms than other arboviruses, such as joint pain, eye commitments and emerging neurological conditions.

Given the poor relationship between the transmission of dengue and entomological indicators, this paper estimated $R_0$ from a SIR model with implicit vector dynamics, following the work of Pandey et al [8].

PAHO established some entomological indices to determine the risk of transmission of dengue[9]. However, it indicated that these indices are not accurate. For example, Focks[10] found a weak relationship between larval indices and production of adult mosquitoes, responsible for transmission. It has also been determined that less than 20% of deposits are responsible for over 80% of *Aedes* adults[11]. Similar results have also been reported by Bowman[12] and Boyer[13]. Particularly the latter suggest that traditional infestation indexes of *Aedes* should not be considered indicators of epidemiological risk. In Medellin, the capital of Antioquia, no relationship between traditional entomological indexes and the incidence of the disease for dengue epidemic registered in 2010 was observed[14].

The estimate of $R_0$ with an implicit vector dynamics uses the data available for epidemiological surveillance, given the limitations of the entomological indicators to establish the potential for transmission or level of risk in the ongoing Zika epidemic.

## Methods

We estimate the basic reproductive number of Zika with daily accounts of cases by municipality, reported to the surveillance system of Antioquia ("SIVIGILA"). We used a simplified SIR model with implicit dynamic vector, as explained in the appendix.

*Data Sources*

Data were obtained from the anonymous database of the public health surveillance system (SIVIGILA) of Antioquia. Case reporting is required by all hospitals or doctors' offices. The case definition and management of the data is based on the guidelines of the Ministry of Health and Social Protection and the National Institute of Health of Colombia (NIH). We include in the analysis all residents in Antioquia, including suspected and confirmed cases by clinical or laboratory. Cases was reported in the first four months of 2016 when the peak of the epidemic occurs[15]. Discarded cases by laboratory confirmation of another disease were excluded.

*Parameter estimation*

The epidemic parameters were estimated using NLREG® version 6.5 (P. Sherrod, TN, USA) by fitting the mathematical expression of individuals recovered per unit time R (t) (see Appendix - equation 18) to the accumulated frequency of daily Zika cases according to the date of onset of symptoms of the index case and successive cases. We verified that the time lapse between the index case and the next case was approximately the extrinsic incubation period of three weeks [19]. The initial guess of the parameters were taken to be $\gamma_h = 0.02$, $R_0$ = 20, s = 100 with 95% confidence interval. Goodness of fit test was used to compare the observed with the estimated data.

As the SIR model and all analysis derived from it deal with retrospective depersonalised routine notification data, ethics approval for the study was not required.

**Results**

Between January 1, 2016 and April 11, 2016 (i.e., week 1 to 15) were reported 1935 cases of Zika in 74 of the 125 municipalities of Antioquia and nine subregions to the Secretary of Health. Of these, 1864 cases were residents in Antioquia (96.3%) in municipalities located below 2200 meters above sea level Table 1.

1841 cases were analyzed, after discarding 23 cases due to other diagnoses. Laboratory confirmation of cases was less than 1% (15/1841). In two municipalities was estimated $R_0$, using data only of clinical or laboratory-confirmed cases, but due the low frequency of confirmation, data of suspected and confirmed cases were used in 11 municipalities and only suspected cases in 7 municipalities Table 1.

Cases were primarily women (67.1%), aged 15 to 44 years old (63.7%) and residents in urban areas (81.8%). These age and gender distributions were similar by municipality Table 1.

We estimated $R_0$ for 20 municipalities that had reported more than 15 cases in the period analyzed. The data of the date of onset of symptoms were available in 99.5% of cases.

Most index cases began showing symptoms in January 2016. Outbreaks were reported during 43 to 98 days (median 79 days) in the municipalities. A median of 38 cumulative cases was reported by municipality, with a minimum of 17 cases in *Sopetrán* and maximum of 347 cases in *Medellin*. Also important was the cumulative frequency of cases in *Apartadó* (n = 311) and *Turbo* (n = 228).

As shown in Figure 2-3, the cumulative cases were growing abruptly in Caceres and Medellin, while in 12 municipalities it stabilized around a fixed number of cases. In the remaining six municipalities, the curve showed a slow growth initially, eventually approaching to a stable value of cumulative number of cases.

Multiple adjusted coefficient of determination was computed and found to be between 96.27% and 99.97%, after obtaining a suitable adjustment of accumulated cases per day and the estimation of individuals recovered per unit time R (t) according to equation 18 (Figure 2-3).

The median $R_0$ was estimated as 1.12 and its upper confidence interval was greater than one in all 16 municipalities wherever estimation was possible.

The greatest potential for transmission occurred in 14 municipalities, having a mean estimated $R_0$ >1 with a maximum of 2.2 (CI 95% 1.54-2.86) in *Nechí* and an extreme value of 56.38 in *Caceres* (Figure 4). These municipalities are part of the *Bajo Cauca* region where the epidemic was in growing stage during the analysis period, as can be seen in Figure 2.

In the subregion of *Valle of Aburrá*, *Medellin* stands with $R_0$ = 22.2, followed by the subregion of *Urabá*, with six municipalities with $R_0$> 1.

Transmission potential was lower in five municipalities as estimated $R_0$ is founded be less than one, but the upper confidence interval included some $R_0$ values greater than one (see Figure 3-4).

## Discussion

Mathematical models have been developed to help understand the epidemiology of vector-borne diseases and support decision making. Some of these models explicitly include dynamic vector population, whereas others do implicitly by considering an effective direct transmission of the disease between human individuals. The model developed in this article belongs to the class of effective direct transmission models. We obtained a proper approximation that provided a simple relationship between the reported cases and the estimated cases from the model without explicitly incorporating the entomological variables and data, following the work of Pandey et al. [8]

Models that explicitly include entomological variables are complex due to their greater number of parameters to be analyzed while is preferred simple models with relatively few parameters and with good consistency with real data, as observed in this analysis.

Our work has the advantage of using the data of the reported frequency of cases available in epidemiological surveillance, to identify regions of transmission of an ongoing epidemic.

This work also highlights the usefulness of detailed, daily and individual data on the onset date of symptoms, to understand the dynamics of transmission in local epidemics. It should be emphasized that estimations of $R_0$ values provide approximations to the local transmission dynamic of Zika. As illustrated here, $R_0$ is used basically to compare the transmission potential between municipalities and accordingly, to identify and suggest effective control measures[20].

This study has some limitations. The data used here are limited because few cases were laboratory confirmed and there is under-diagnosis or underreporting of cases that could lead to sub-estimation of $R_0$. Additionally, the presence of asymptomatic individuals so far estimated at 80% in other contexts[21] [3] could affect estimation $R_0$ in our model that assumes that we are counting all infectious cases.

The comparisons of our estimations of $R_0$ for Zika with the reported values in the literature is difficult to carry out as different models are used by different studies.

Towers et al [22], used a SEIR / SEI model for estimating $R_0$. They used data from the daily incidence of suspected Zika cases reported in Barranquilla from October to November 2015, to estimate the exponential growth rate and they also used vector parameters values from the literature. According to authors, $R_0$ estimated was 4.4 (CI 95% 3. 0 - 6.2). These authors used a similar expression to equation 17 and some parameters were obtained assuming that the initial phase of the epidemic generates an exponential curve. Since only the data from the beginning of the epidemic were used, the exponential growth may overestimate the prevalence and therefore the reproduction number. In our model we use the equation 16, a shortened form of the equation 17 and 18 representing the full course of the epidemic and not just their initial phase; therefore our method tends to underestimate the prevalence and the reproduction number.

Other authors have analyzed data from the initial phase of the epidemic based on the exponential growth rate, using phenomenological models.

Majumder, et al.[23], used exponential smoothing models as well as Incidence Decay and Exponential Adjustment (IDEA) model to estimate $R_0$ for Colombia and other countries, using the published digital data HealthMap and the NIH. They estimated an average $R_0$ of 4.82 (range 2.34-8.32).

Nishiura, et al.[24], estimated $R_0$ from Colombia, using maximum likelihood methods. They assumed that the confirmed cases from week 35 of 2015 had an exponential growth. The estimate of $R_0$ range was from 3.0 to 6.6.

Chowell et al, [15] estimated $R_0$ by calculating the polynomial growth profile of the initial trajectory of the epidemic, with data from the daily incidence of Antioquia globally. Assuming a gamma distributed generation interval (average: 14 days S.D 2) they estimated $R_0$ of 10.3 (CI95%: 8.3 -12.4) in the first generation of Zika to 2.2 (CI95%: 1.9 -2.8) in the second generation. When the exponential distributed generation interval was assumed, $R_0$ was estimated from 2.8 (CI95%: 2.4 - 3.1) to 1.8 (CI95%: 1.7- 2.0) for the two generations, respectively. The latter values were lower than those estimated by Nishiura et al, 2016 who used weekly data from Colombia, but are closer to those obtained in this work, for each of the municipalities, despite using different models.

The age and sex distribution of cases of Antioquia, are similar to those reported in Colombia[5]. As has been documented, the tendency of women to consult is related to concerns about the risk of congenital malformations after infection of pregnant women. The occurrence of cases residents in urban areas and population aged 15 to 44 years shows possibly the difference in exposure or susceptibility of these populations.

The results show a greater transmission of Zika in municipalities of *Bajo Cauca, Uraba* regions and *Medellin*. The greatest $R_0$ in the municipalities of these regions is in line with observed and indicates the need to strengthen the capacity of detection and case management and strengthening the integrated vector control in such places.

## Appendix

### From the Vector SIR Model to the simple SIR model

The standard SIR vector model is given by the following system of nonlinear and coupled differential equations where the equations (1), (2) and (3) describe the human population classified as susceptible ($S_h$), infected ($i_h$) and removed ($R_h$); and equations (4) and (5) describe the mosquito population classified as susceptible ($S_m$) and infected ($i_m$) [16]. We assumed that the mortality, migration and natality were negligible because the short time scale of the epidemic occurrence.

$$\frac{dS_h(t)}{dt} = -\beta_{mh}\, S_h(t)\, i_m(t) \tag{1}$$

$$\frac{di_h(t)}{dt} = \beta_{mh}\, S_h(t)\, i_m(t) - \gamma_h\, i_h(t) \tag{2}$$

$$\frac{dR_h\ (t)}{dt} = \gamma_h\, i_h(t) \tag{3}$$

$$\frac{dS_m(t)}{dt} = \mu_m N_m - \beta_{hm}\, S_m(t)\, i_h(t) - \mu_m\, S_m(t) \tag{4}$$

$$\frac{di_m}{dt}(t) = \beta_{hm}\, S_m(t)\, i_h(t) - \mu_m\, i_m(t) \tag{5}$$

where

$\beta_{hm} \approx b\widetilde{\beta_{hm}}$

$\beta_{mh}(\approx b\widetilde{\beta_{mh}}\frac{N_m}{N_h}$

And the model parameters are as follows:

$b$: bitting rate

$\beta_{h,m}$ : probability of effective transmission from human to mosquito by bite.

$\beta_{m,h}$ : probability of effective transmission from mosquito to human by bite.

$\gamma_h$ : per capita recovery rate of Zika infection in humans

$\mu_m$ : per capita mortality rate in mosquitoes.

$N_h$: total human population

$N_m$: total mosquito population.

We assume that the densities of susceptible mosquitoes and infected mosquitoes are constant over time and obtain a simplified SIR model as below. That is,

$$i_m(t) = i_m \tag{6}$$

$$S_m(t) = S_m \tag{7}$$

Substituting (6) and (7) in (5) we obtain:

$$0 = \beta_{hm}\, S_m\, i_h(t) - \mu_m\, i_m \tag{8}$$

From (8), it follows that:

$$i_m = \frac{\beta_{hm}\, S_m\, i_h(t)}{\mu_m} \tag{9}$$

Using (6) and (9) in (1) we obtain:

$$\frac{dS_h}{dt}(t) = -\frac{\beta_{mh}\, S_h(t)\, \beta_{hm}\, S_m\, i_h(t)}{\mu_m} \tag{10}$$

Using (6) and (9) in (2) we obtain:

$$\frac{di_h}{dt}(t) = \frac{\beta_{mh}\, S_h(t)\, \beta_{hm}\, S_m\, i_h(t)}{\mu_m} - \gamma_h\, i_h(t) \tag{11}$$

Since $S_m$ is assumed constant, we define average effective infectivity as in Pandey et al. (2013)[8]

$$\beta = \frac{\beta_{mh}\, \beta_{hm}\, s_m}{\mu_m} \tag{12}$$

Hence the equations (10) and (11) can be rewritten as:

$$\frac{d}{dt} S_h(t) = -\beta\, S_h(t)\, i_h(t) \tag{13}$$

$$\frac{d}{dt} i_h(t) = \beta\, S_h(t)\, i_h(t) - \gamma_h\, i_h(t) \tag{14}$$

Rewriting the equation (3)

$$\frac{d}{dt} R_h(t) = \gamma_h R_h(t) \ldots\ldots\ldots\ldots\ldots\ldots\ldots\ldots\ldots\ldots\ldots\ldots\ldots\ldots\ldots \tag{15}$$

It shows that the human population satisfies a simple SIR model given by equations (13), (14) y (15).

For the *effective* SIR model ((13), (14) and (15)) the basic reproduction number $R_0$ is given by:

$${R_0}^2 = \frac{S_h(0)\, \beta}{\gamma_h} \tag{16}$$

Substituting (12) in (16), we obtain an $R_0$ which is similar to the reproduction number of the classic Ross-Macdonald model

$${R_0}^2 = \frac{S_h(0)\, \beta_{mh}\, \beta_{hm}\, s_m}{\mu_m \gamma_h} \tag{17}$$

it is to say

$$R_0{}^2 \left[= \frac{S_h(0)\,\beta_{mh}}{\gamma_h} \cdot \frac{\beta_{hm}\,S_m}{\mu_m}\right]$$

where $\frac{S_h(0)\beta_{mh}}{\gamma_h}$ is the average number of new infected humans generated by an infectious mosquito, and $\frac{S_m\beta_{hm}}{\mu_m}$ represents the average number of new infected mosquitoes generated by a single infected human.

**An approximate solution for the effective SIR model and obtaining $R_0$**

The SIR model (13), (14) and (15) can be solved approximately: [8 16, 17]

$$R(t) = \frac{\rho^2\left(\frac{s}{\rho} - 1 - \alpha \tanh(-0.5\,\alpha\,\gamma_h\,t + \phi)\right)}{s} \tag{18}$$

where $s = S_h(0)$; and

$$\alpha = \sqrt{\left(\frac{s}{\rho} - 1\right)^2 + \frac{2\,s}{\rho^2}} \tag{19}$$

$$\phi = \frac{1}{2}\ln\left(\frac{\alpha\,\rho + s - \rho}{\alpha\,\rho - s + \rho}\right) \tag{20}$$

$$\rho = \frac{\gamma_h}{\beta} \tag{20A}$$

**Algorithm to estimate the basic reproductive number** using NLREG ® version 6.5 (P. Sherrod, TN, USA)

*Title "Cumulated cases of Zika 2015-2016";*

*Variables Day, CC;*

```
Parameter gamma =0.02;
Parameter R =20;
Parameter s =100;
Double rho, alpha, phi;
rho=s/R;
alpha = ((s/rho-1)^2+2*s/rho^2)^(1/2);
phi =1/2*ln((alpha*rho+s-rho)/(alpha*rho-s+rho));
Function CC =rho^2/s*(s/rho-1-alpha*tanh(-.5*alpha*gamma*Día+phi));
Plot xvar=Day, xlabel="Time (Day)", ylabel="Cumulated cases";
Plot;
rplot;
ITERATIONS 100;
CONFIDENCE 95;
Data;
```

**Funding Statement**

This work was partially funded by University of Antioquia. The funder had no role in study design, data collection and analysis, decision to publish, or preparation of the manuscript.

**Competing Interest Statement**

The authors have declared that no competing interests exist

**Acknowledge**

The authors gratefully acknowledge useful comments on earlier draft by Gerardo Chowell (Arizona State University) and Francisco J Diaz (University of Antioquia)

Table. Zika cases of residents in Antioquia- Colombia, 2016 by subregion and municipality

| Sub region/ Municipality | Meters above sea level | Type of cases | Dates of index case and last case reported | Ratio urban/rural | Age – number(percent) | | |
|---|---|---|---|---|---|---|---|
| | | | | | <15 | 15-44 | 45 + |
| BAJO CAUCA | | | | | | | |
| Cáceres | 200 | Suspect | 03/01/2016 15/02/2016 | 1.0 | 14 (31.1) | 21 (46.6) | 10 (22.2) |
| Nechí | 30 | Suspect | 30/01/2016 20/03/2016 | 4.8 | 9 (31.3) | 12 (41.4) | 8 (17.8) |
| Caucasia | 150 | Suspect & Confirmed | 10/01/2016 05/04/2016 | 13.8 | 11 (18.6) | 37 (66.1) | 11 (18.6) |
| Zaragoza | 150 | Suspect | 29/01/2016 07/04/2016 | 7.2 | 7 (14.3) | 34 (69.4) | 8 (16.3) |
| VALLE ABURRÁ | | | | | | | |
| Medellín | 1538 | Suspect & Confirmed | 12/09/2015 09/04/2016 | 13.0 | 35 (10.0) | 231 (65.8) | 85 (24.2) |
| Bello | 1450 | Suspect & Confirmed | 11/01/2016 01/04/2016 | 7.0 | 10 (31.3) | 15 (46.8) | 7 (21.9) |
| Envigado | 1575 | Suspect & Confirmed | 01/01/2016 03/04/2016 | 6.0 | 2 (9.5) | 12 (57.1) | 7 (33.3) |
| Itagüí | 1550 | Suspect & Confirmed | 13/01/2016 03/04/2016 | 3.4 | 3 (8.1) | 31 (83.8) | 3 (8.1) |
| URABÁ | | | | | | | |
| Chigorodó | 34 | Suspect & Confirmed | 02/01/2016 06/04/2016 | 17,2 | 31 (18,9) | 105 (64,1) | 28 (17,0) |
| San Pedro de Urabá | 200 | Suspect & Confirmed | 06/01/2016 30/03/2016 | 2.3 | 4 (17.4) | 9 (39.1) | 10 (43.5) |
| Carepa | 28 | Suspect & Confirmed | 07/01/2016 06/04/2016 | 4.1 | 14 (13.7) | 75 (73.5) | 13 (12.7) |
| Turbo | 2 | Suspect & Confirmed | 02/01/2016 02/04/2016 | 4.4 | 44 (19.3) | 153 (67.1) | 31 (13.6) |
| Necoclí | 8 | Suspect & Confirmed | 25/01/2016 07/04/2016 | 1.0 | 5 (13.9) | 22 (61.1) | 9 (25.0) |
| Apartadó | 30 | Suspect & Confirmed | 02/01/2016 10/04/2016 | 7.1 | 64 (20.3) | 196 (62.2) | 55 (17.5) |
| Mutatá | 75 | Suspect & Confirmed | 26/01/2016 03/04/2016 | 1.7 | 6 (31.6) | 11 (57.9) | 2 (10.5) |
| NORDESTE | | | | | | | |
| Remedios | 700 | Suspect | 16/02/2016 04/04/2016 | 3.5 | 3 (16.7) | 15 (83.3) | 0 (0.0) |
| MAGDALENA MEDIO | | | | | | | |

| Sub region/ Municipality | Meters above sea level | Type of cases | Dates of index case and last case reported | Ratio urban/rural | Age – number(percent) | | |
|---|---|---|---|---|---|---|---|
| | | | | | <15 | 15-44 | 45 + |
| Puerto Berrío | 125 | Suspect | 31/01/2016 02/04/2016 | 3.7 | 10 (16.4) | 34 (55.7) | 17 (27.9) |
| Puerto Triunfo | 150 | Suspect | 06/01/2016 02/04/2016 | 0.6 | 7 (25.9) | 17 (63.0) | 3 (11.1) |
| OCCIDENTE | | | | | | | |
| Sopetrán | 750 | Suspect | 15/01/2016 01/04/2016 | 1.8 | 0 (0.0) | 8 (47.0) | 9 (52.9) |
| ORIENTE | | | | | | | |
| Rionegro | 2125 | Suspect & Confirmed | 26/12/2015 02/04/2016 | 1.3 | 0 (0.0) | 15 (83.3) | 3 (16.7) |

Figure 1. Sub regions of Antioquia, Colombia

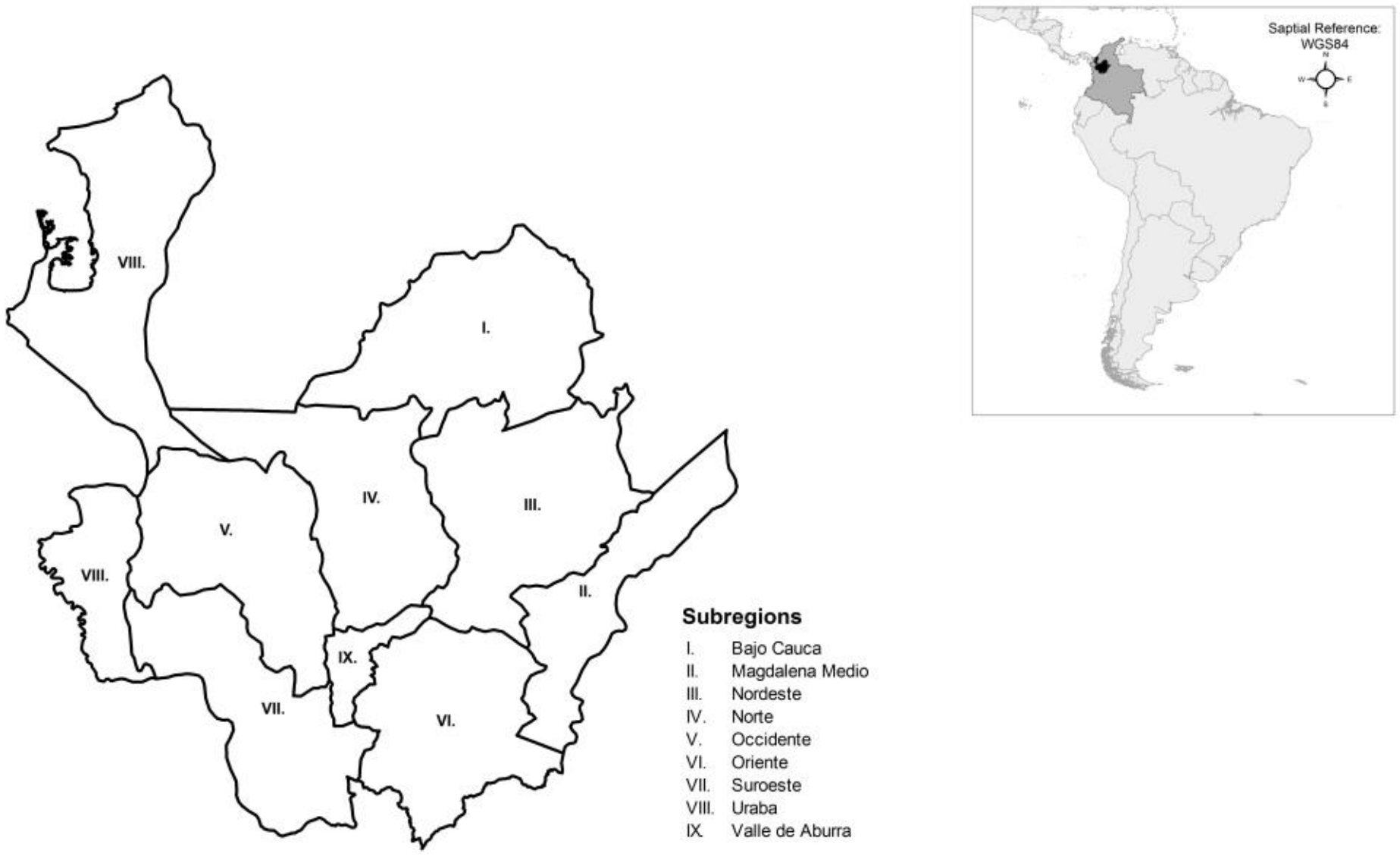

Figure 2. High Potential transmission ($R_0$>1) of Zika by municipality and subregion of Antioquia- Colombia, 2016.

Bajo Cauca Subregion

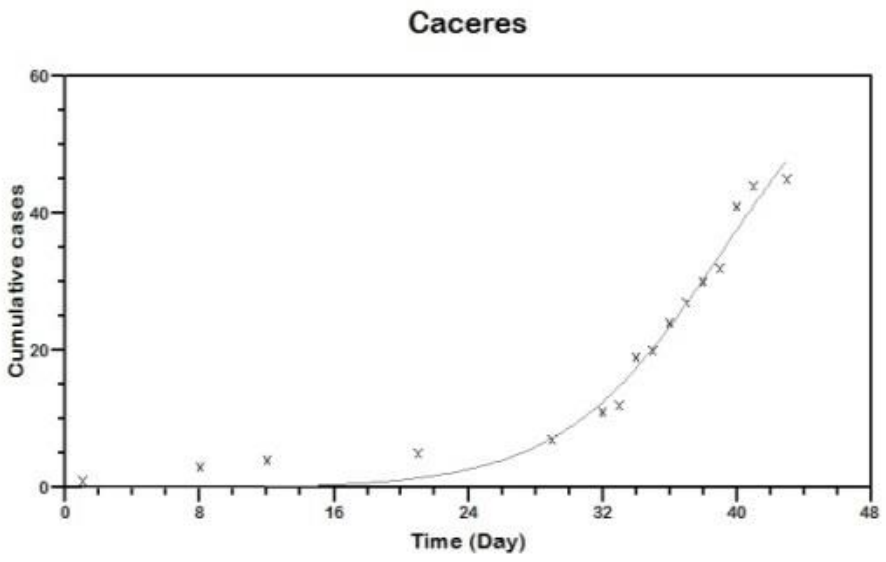

Ro= 56.38 (CI95% n.e.)
Adj.Coef.Det.=97.24%

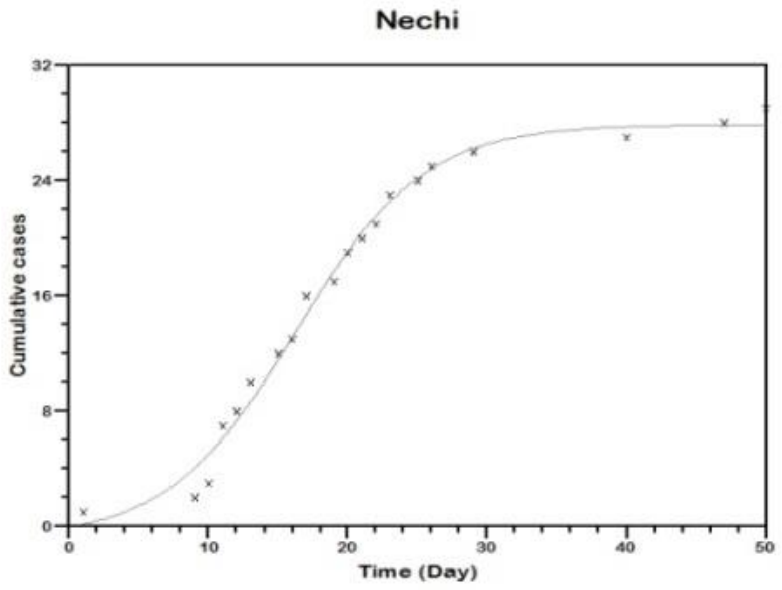

Ro= 2.2 (CI95% 1.54-2.86)

Adj.Coef.Det.=98.78%

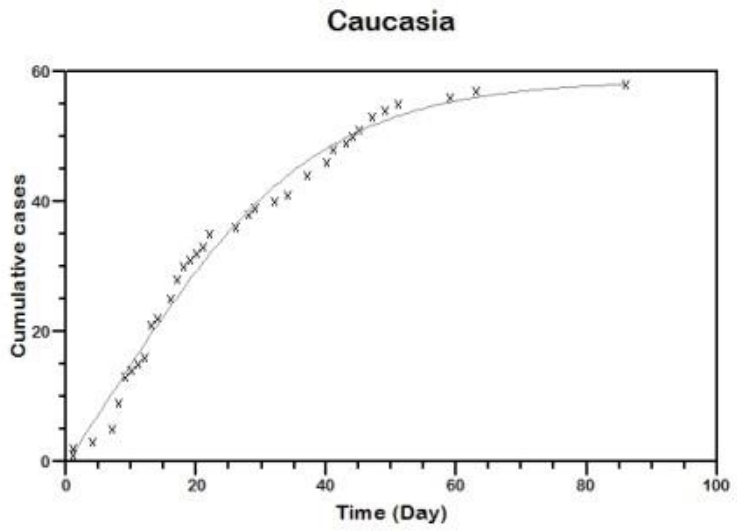

Ro= 1.009 (CI95% 0.99-1.02)
Adj.Coef.Det.=98.54%

Urabá Subregion

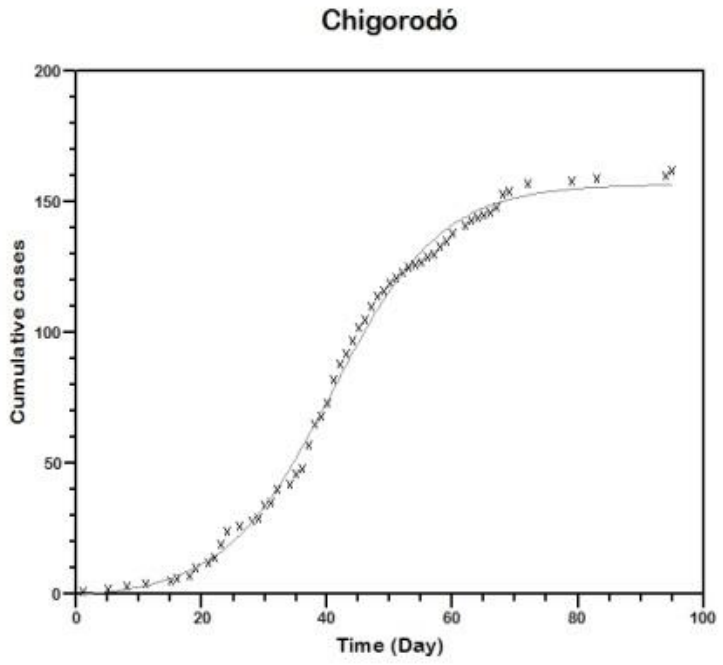

Ro= 1.53 (CI95% 1.71-1.89)
Adj.Coef.Det.=99.59%

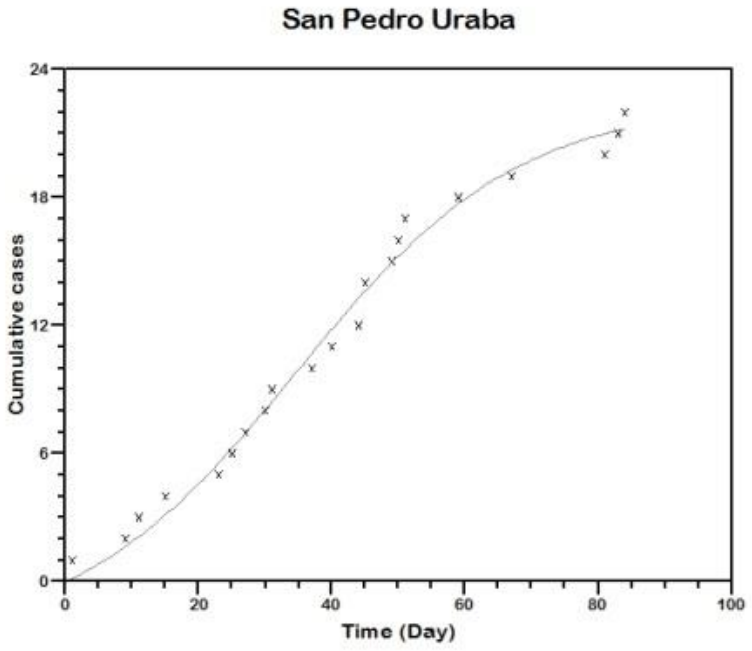

Ro= 1.31 (CI95% 1.10-1.52)
Adj.Coef.Det.==98.69%

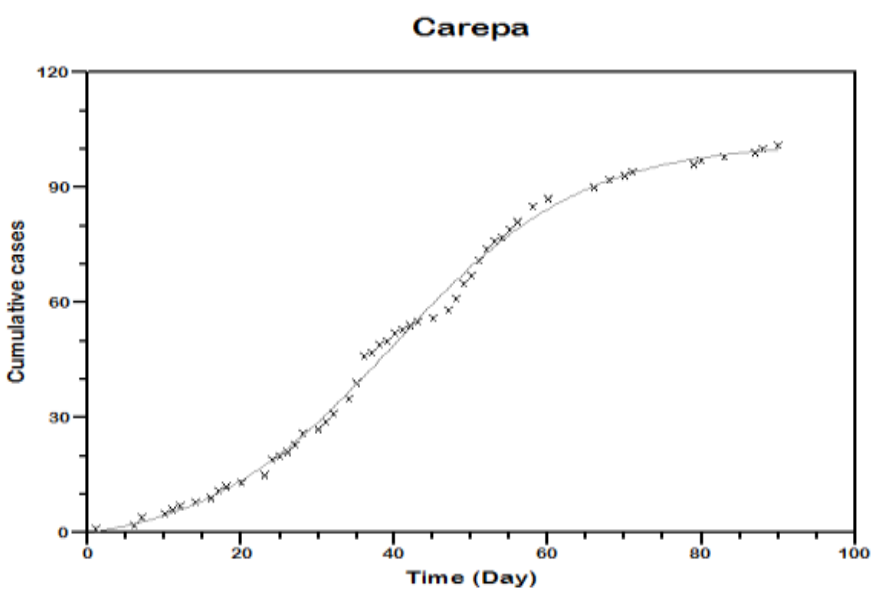

$R_0$= 1.23 (CI95% 1.18-1.28)
Adj.Coef.Det.=99.6%

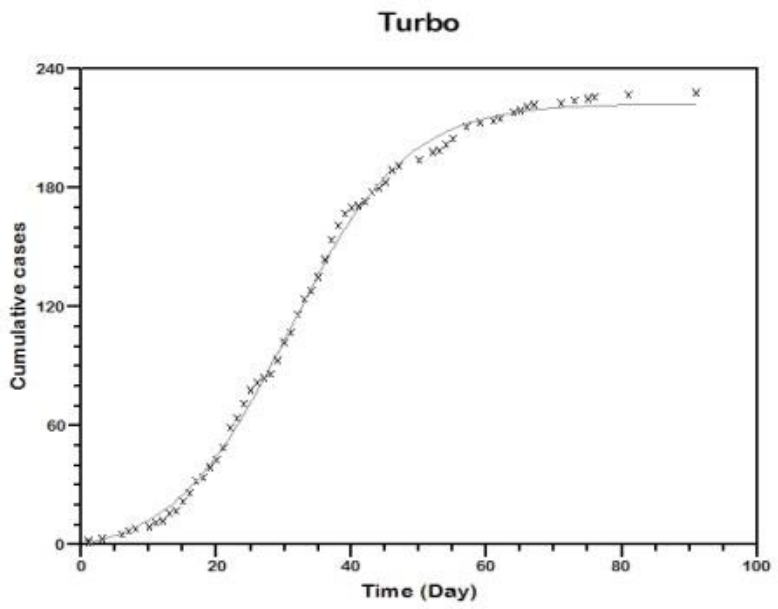

$R_0$= 1.15(CI95% 1.12-1.17)
Adj.Coef.Det.=99.7%

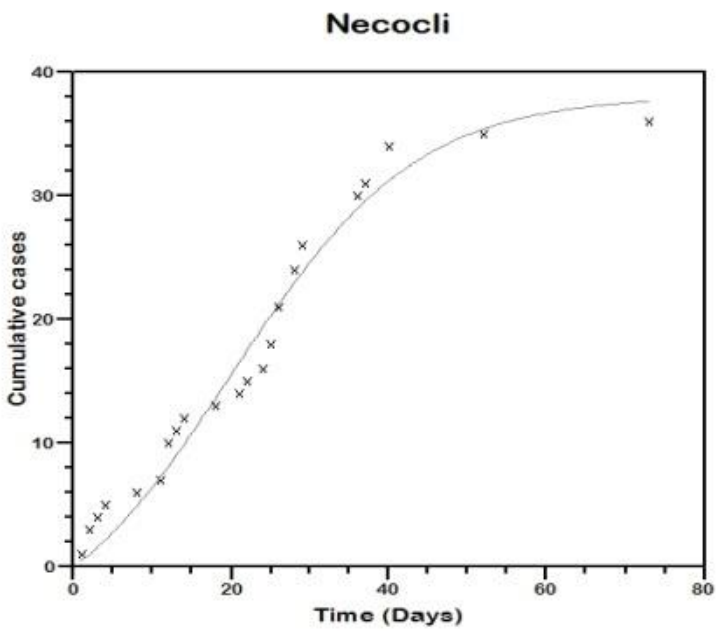

$R_0$= 1.12 (CI95% 0.99-1.24)
Adj.Coef.Det.=96.64%

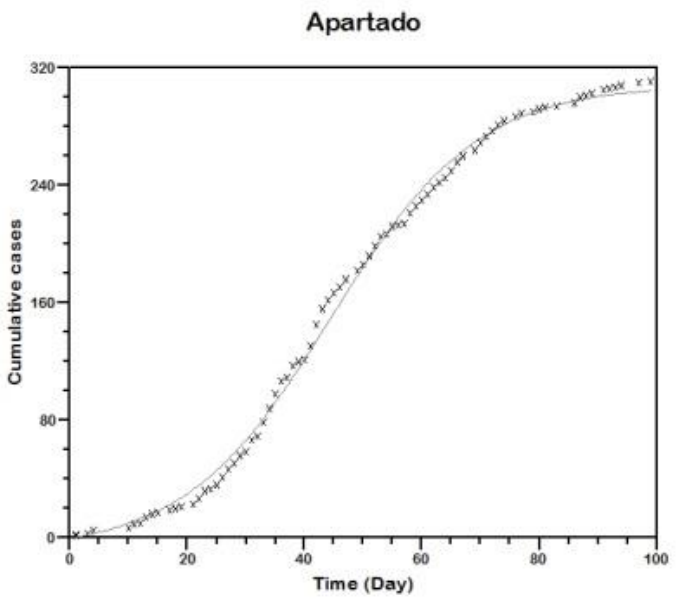

$R_0$= 1.11 (CI95% 1.09-1.13)
Adj.Coef.Det.=99.67%

Valle de Aburrá Subregion

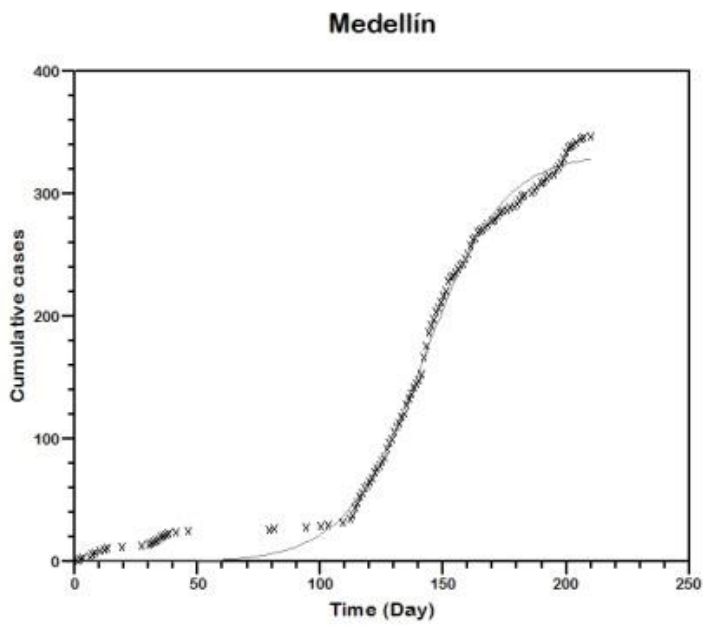

$R_0$= 22.2 (CI95% n.e.) Adj.Coef.Det.=99.3%

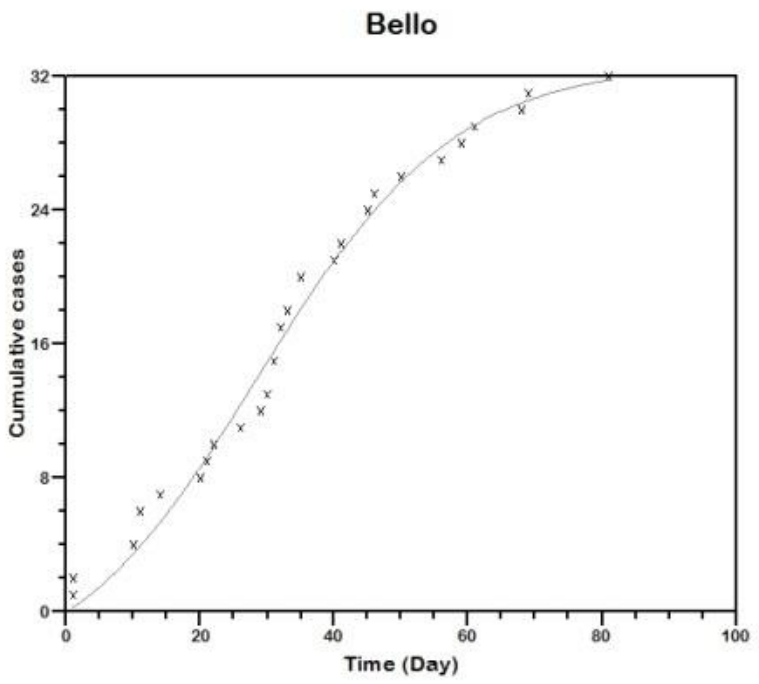

$R_0$= 1.18 (CI95% 1.06-1.3)
Adj.Coef.Det.=98.58%

a) Other Subregion

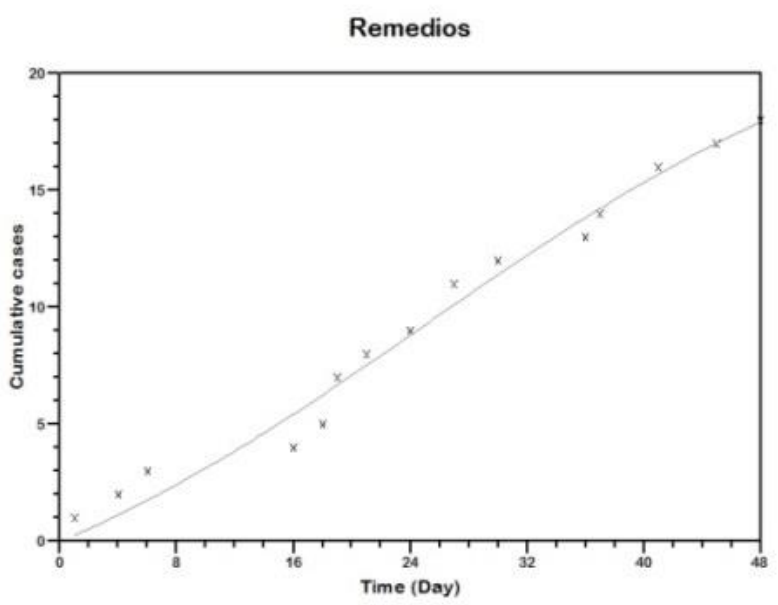

$R_0$= 1.13 (CI95% 0.84-1.42)
Adj.Coef.Det.=97.67%

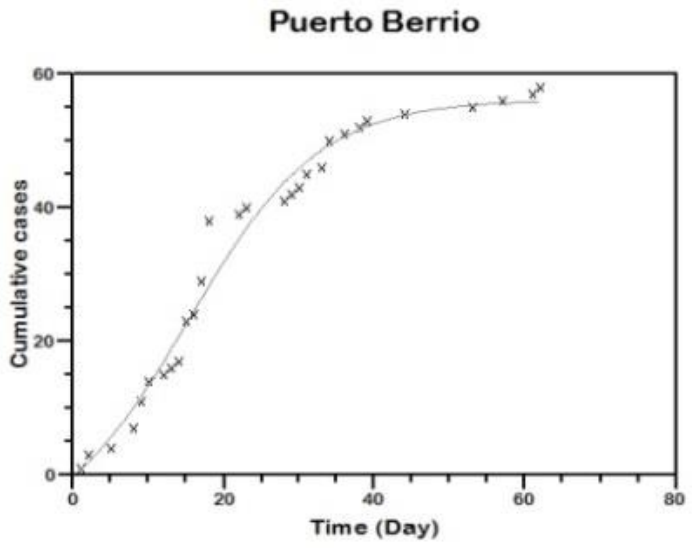

$R_0$= 1.08 (CI95% 1.02-1.15)
Adj.Coef.Det.=97.7%

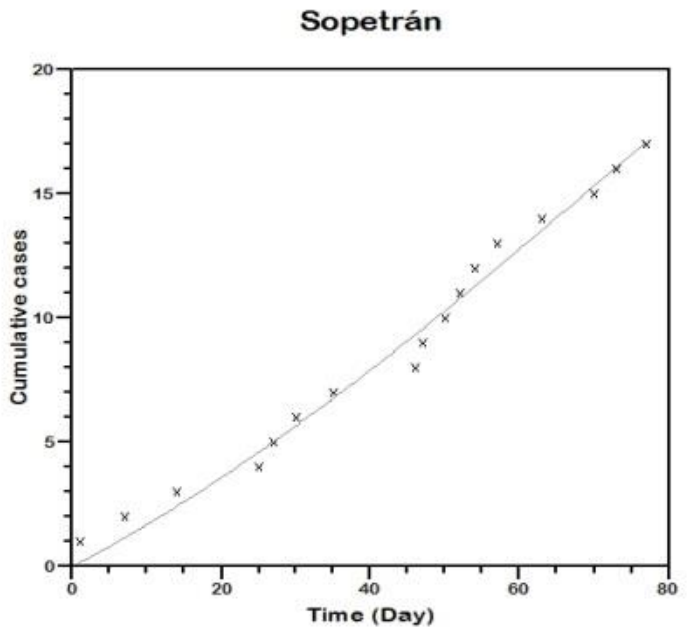

$R_0$= 1.07 (CI95% 0.8-1.26)
Adj.Coef.Det.=98.24%

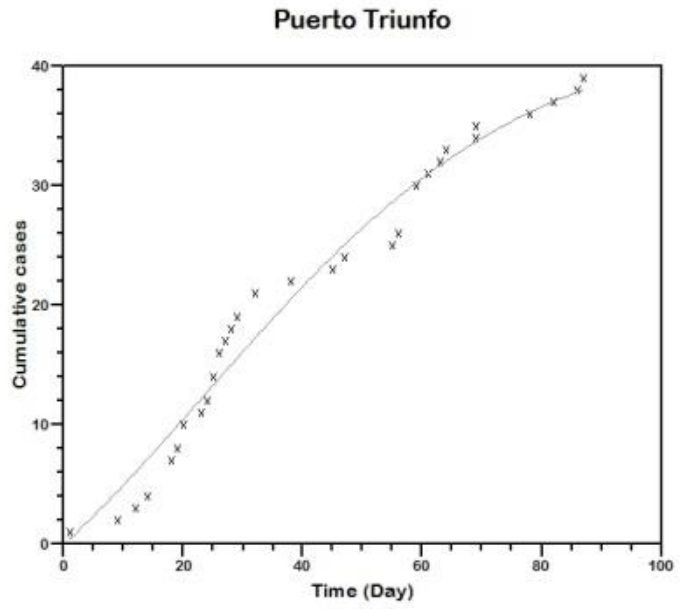

$R_0$= 1.03 (CI95% 0.97-1.09)
Adj.Coef.Det.=97.0%

Figure 3. Less potential transmission ($R_0$<1) of Zika by municipality and subregion of Antioquia- Colombia, 2016.

a) Urabá Subregion

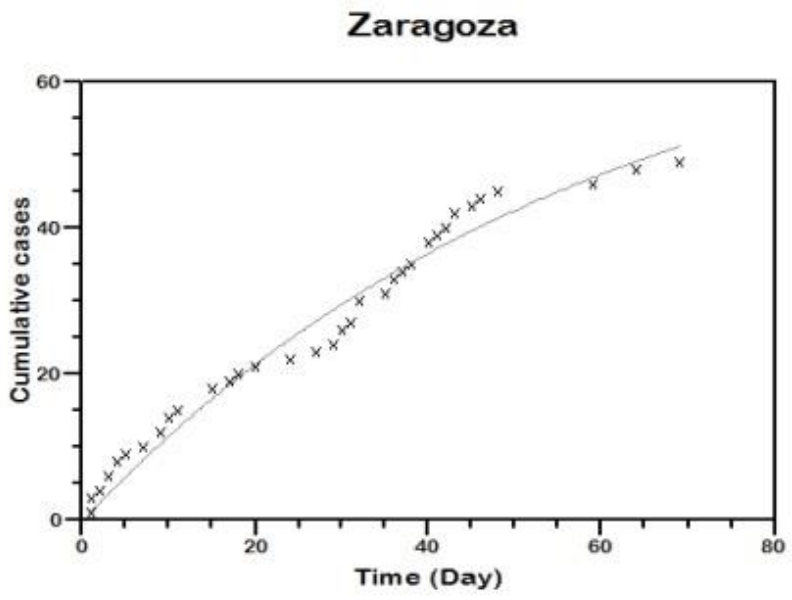

$R_0$= 0.98 (CI95% n.e)
Adj.Coef.Det.=96.96%

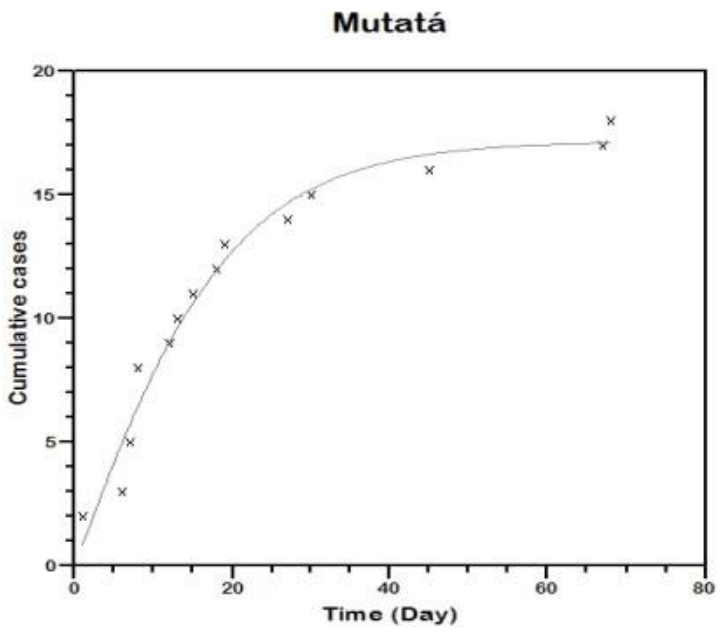

$R_0$= 0.98 (CI95% 0.88-1.08)
Adj.Coef.Det.=96.27%

b) Valle de Aburrá Subregion

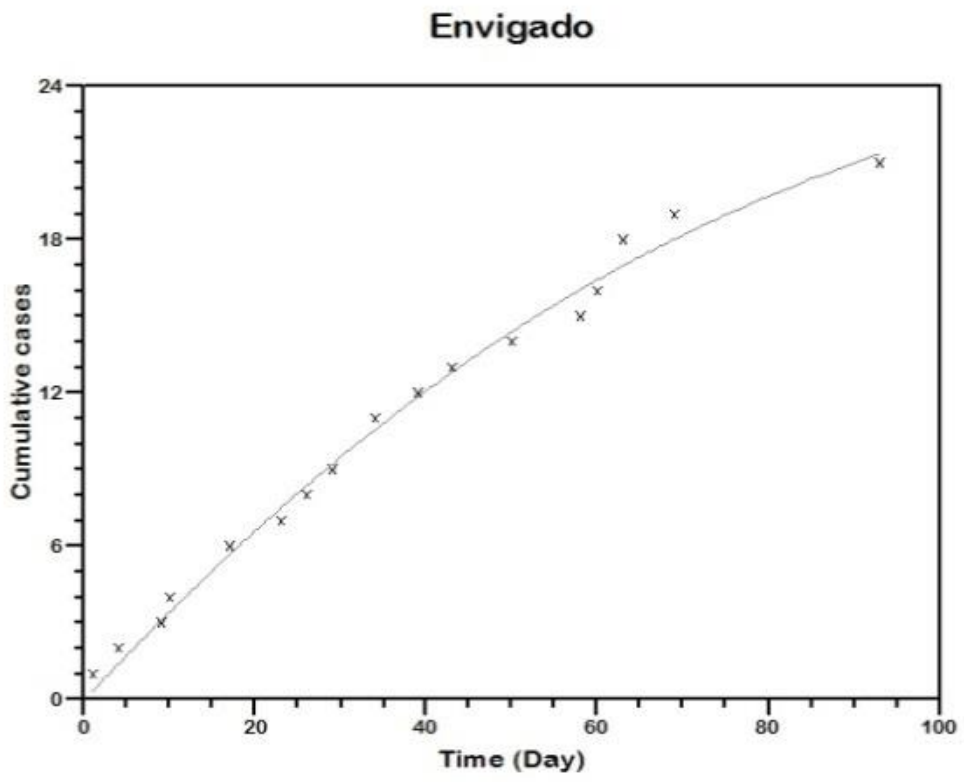

$R_0$= 0.98 (CI95% 0.93-1.02)
Adj.Coef.Det.=98.98%

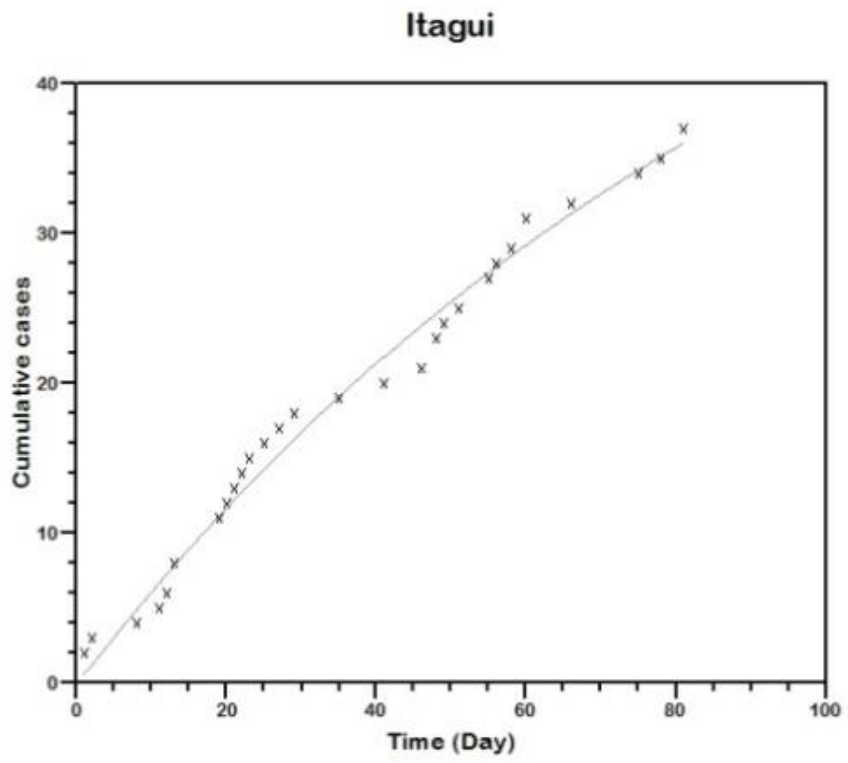

$R_0$= 0.98 (CI95% n.e)
Adj.Coef.Det.=98.31%

c) Oriente Subregion

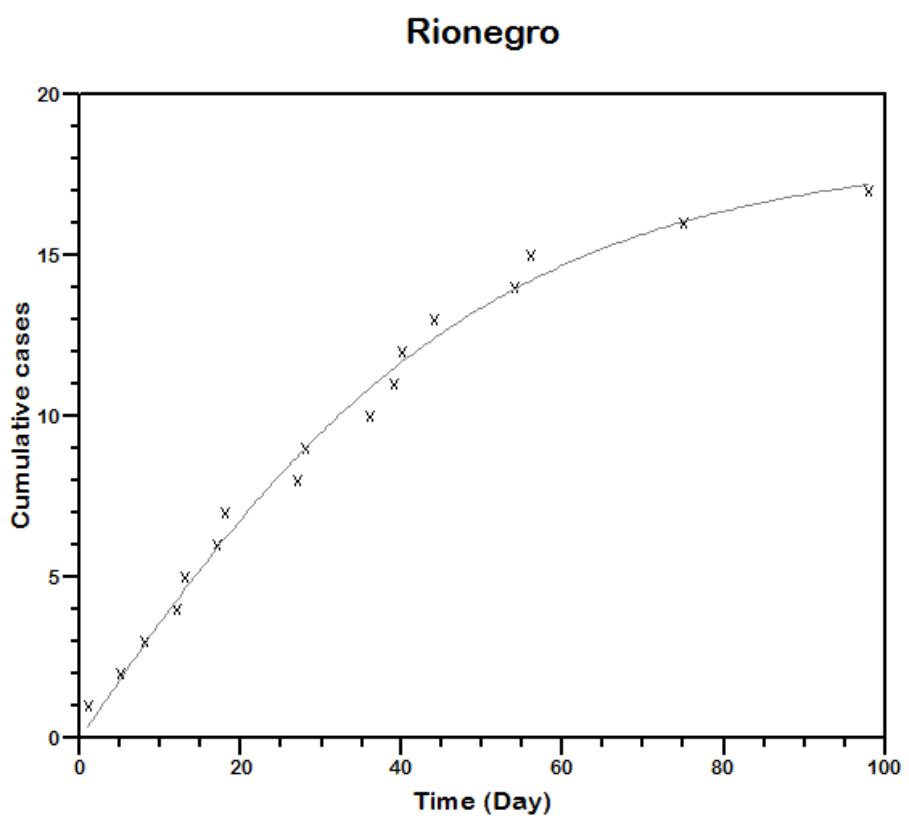

$R_0$= 0.98 (CI95% 0.92-1.03)
Adj.Coef.Det.=98.9%

Figure 4. Reproductive number of Zika in municipalities of Antioquia, Colombia

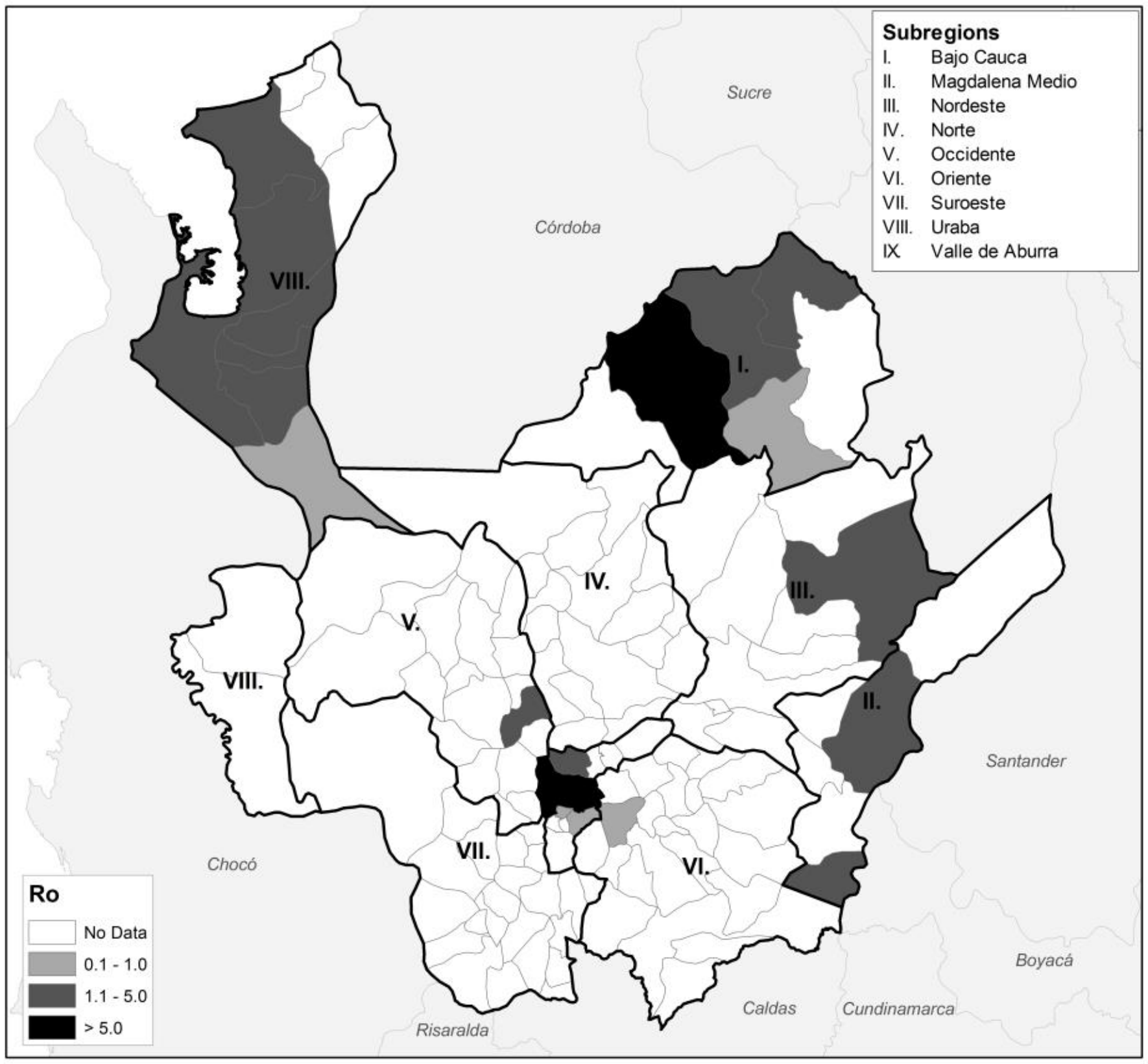